\documentclass[aps,pre,twocolumn,superscriptaddress,floatfix]{revtex4-2}
\usepackage[utf8]{inputenc}
\usepackage{amsmath,amssymb,graphicx,bm}
\usepackage[caption=false]{subfig}

\makeatletter
\long\def\@makecaption#1#2{%
  \vskip\abovecaptionskip
  \small
  \sbox\@tempboxa{#1: #2}%
  \ifdim \wd\@tempboxa > \hsize
    \parbox{\hsize}{\justifying #1: #2}%
  \else
    \parbox{\hsize}{\justifying #1: #2}%
  \fi
  \vskip\belowcaptionskip}
\makeatother
\usepackage{ragged2e} 

\usepackage[colorlinks=true,linkcolor=blue,citecolor=blue,urlcolor=blue]{hyperref}

\begin{document}

\title{Chaos in Nonequilibrium Two-Temperature ($T_x$, $T_y$) Nos\'e--Hoover Cell Models}

\author{Hesam Arabzadeh}
\affiliation{Department of Chemistry, University of Missouri, Columbia, Missouri 65211-7600, USA}

\author{Carol Griswold Hoover}
\affiliation{Ruby Valley Research Institute, Unit 109, 2870 Ruby Vista Drive, Elko, Nevada 89801, USA}

\author{William Graham Hoover}
\affiliation{Ruby Valley Research Institute, Unit 109, 2870 Ruby Vista Drive, Elko, Nevada 89801, USA}

\author{Brad Lee Holian}
\affiliation{Theoretical Division, Los Alamos National Laboratory, Los Alamos, New Mexico 87545, USA}

\begin{abstract}
\textbf{Abstract}\\
We revisit a two-temperature Nos\'e--Hoover wanderer particle embedded in a two-dimensional periodic $2\times 2$ cell with four smooth repulsive corners at $(x,y) = (\pm 1, \pm 1)$ to explore chaos with anisotropic thermostatting. The model employs separate thermostats in the $x$ and $y$ directions, enabling controlled deviations from equilibrium. By integrating the full six-dimensional equations of motion and computing the complete Lyapunov spectrum, we confirm chaos and quantify phase-space contraction with high numerical precision. The total contraction rate, interpreted as entropy production, increases nonlinearly with the thermostat anisotropy, deviating from the quadratic dependence expected from linear-response theory, $\Lambda \propto \delta^2$. We compare two fits for $\Lambda$ as a function of $\delta = 0.5 - T_y$: (1) a power law,  $\Lambda \propto \delta^{2.44}$ and (2) a quadratic-plus-quartic expansion. While the former captures low-driving behavior slightly better, the latter more accurately describes the strongly driven regime and remains consistent with linear-response theory near equilibrium. An empirical linear relation between dissipation and phase-space dimensionality loss is also identified, $\Lambda \approx (D_{KY} - 6)/3$, where $D_{\rm KY}$ is the approximate Kaplan--Yorke dimension. Momentum statistics show significant non-Gaussian behavior under strong driving. Despite its dissipative nature, the model remains strictly time-reversible, offering a pedagogically rich example of microscopic reversibility coexisting with macroscopic entropy production.
\end{abstract}

\maketitle

\section{Introduction}

The accurate and efficient simulation of nonequilibrium phenomena often relies on deterministic thermostats, algorithms designed to enforce a desired temperature in molecular or particle dynamics without introducing random forces. Among these, the Nos\'e--Hoover thermostat~\cite{hoover1985canonical,nose1984molecular,hoover2024canonical} is notable for its ability to generate canonical-ensemble dynamics through extended equations of motion. Over the past several decades, Nos\'e--Hoover methods have been employed to study a wide range of systems, from simple fluids to complex materials~\cite{evans2008statistical}.

Typically~\cite{martyna1992nose}, a single thermostat variable controls the temperature of the entire system, ensuring $\langle p^2 \rangle = mkT$ in two dimensions. However, a two-temperature variant of the Nos\'e--Hoover approach introduced some time ago~\cite{hoover1991heat} has received little attention, possibly due to its complex behavior and the lack of immediate applications. We do not propose a specific application here but note that modern advances could render this scheme highly relevant in anisotropic or directionally driven systems. We choose the minimal one-particle model to isolate the thermostat’s role, unobscured by many-body complications. In this two-temperature model, separate variables regulate the motion along different coordinates, allowing $T_x \neq T_y$  in two dimensions (or $T_x \neq T_y \neq T_z$ in three dimensions). This naturally produces a nonequilibrium heat flow among the $x$, $y$ (or $x$, $y$, $z$) directions. In practical terms, directionally distinct temperature controls may arise in layered or rod‐like materials with anisotropic thermal conductivity, or in microfluidic channels subject to asymmetric heating or spatially non‑uniform heating~\cite{squires2005microfluidics,cahill2014nanoscale}. Although this minimal model does not explicitly incorporate fluid flow, it can nevertheless illustrate how temperature anisotropy influences chaotic dynamics. Despite its apparent simplicity, the model poses intriguing questions: How does the phase space evolve under multiple target temperatures? Does the system become chaotic, forming a fractal attractor? What is the net dissipation or entropy production?

To explore these questions, we apply a two-thermostat Nos\'e--Hoover method~\cite{hoover1991heat} with a repulsive containment potential~\cite{hoover2015comparison} in the four quarters of a periodic two-dimensional $2\times 2$ box.We then compute the full six-dimensional Lyapunov spectrum and confirm the presence of chaotic trajectories. Next, we demonstrate that the driven system occupies a lower‐dimensional attractor. We estimate the fractal dimension using the approximate Kaplan–Yorke formula. We also discuss how the attractor dimension and entropy production (dissipation) grow with the temperature difference $|T_x - T_y|$. Furthermore, although the system is non‐Hamiltonian and dissipative, we emphasize that it remains time‐reversible as described in~\cite{evans1993probability,hoover2024canonical}. Reversing the momenta $(p_x, p_y)$ and thermostat variables $(\zeta_x, \zeta_y)$ reproduces the trajectory backward in phase space. A summary of our findings is provided in the concluding section.

\section{Model and Numerical Methods}

\subsection{Equations of Motion}

We consider a single wanderer particle in the $(x,y)$ plane, with momenta $(p_x, p_y)$. Extending the conventional Nos\'e--Hoover approach, we introduce two thermostat variables $(\zeta_x, \zeta_y)$, each controlling the kinetic energy in one coordinate direction in our two-dimensional model. The phase space is thus six-dimensional: $(x, y, p_x, p_y, \zeta_x, \zeta_y)$. 

\subsubsection{Non-Hamiltonian (Two-Thermostats)}
\label{sec:model}
The extended Nos\'e--Hoover equations of motion are
\begin{align}
m\dot{x} &= p_x, \label{eq:eom1}\\
m\dot{y} &= p_y, \label{eq:eom2} \\
\dot{p}_x &= -\frac{\partial \phi}{\partial x} - \zeta_x p_x, \label{eq:eom3} \\
\dot{p}_y &= -\frac{\partial \phi}{\partial y} - \zeta_y p_y,  \label{eq:eom4}\\
\dot{\zeta}_x &= \frac{\left[p_x^2/m - kT_x\right]}{kT_x \tau^2_x}, \label{eq:eom5} \\
\dot{\zeta}_y &= \frac{\left[p_y^2/m - kT_y\right]}{kT_y \tau^2_y}.  \label{eq:eom6}
\end{align}
For convenience, the wanderer mass $m$, the Boltzmann constant $k$, and the thermostat relaxation times $\tau_x$, $\tau_y$ are all set to unity. All variables and numbers are expressed in reduced units throughout this work. We use a smooth repulsive corner potential of the form
\begin{equation}
\phi(r) = \sum_{i=1}^{4}  ~[1 - |r_w - r_i|^2]^4 ~ ~~ \text{for} ~~ ~ |r_w - r_i| < 1,
\end{equation}
where $r_w = (x,y)$ is the position vector of the wanderer particle, and the four $r_i$ are the coordinates of the four fixed scatterers located at the cell corners $(\pm 1, \pm 1)$. The squared distance $ |r_w - r_i|^2$ is used to evaluate the corner potential. All coordinates are measured from the center of the $2\times 2$ cell, spanning $[-1,1]$ in both $x$ and $y$. The polynomial inside the bracket $1 - |r_w - r_i|^2$ combines linear and quadratic terms in a non-monotonic way, peaking at intermediate values of $ |r_w - r_i|^2$ and vanishing smoothly near the boundaries of the interaction range. Raising this expression to the fourth power sharpens the repulsion near the corners while keeping the potential fully differentiable and finite everywhere. This form was chosen in reference ~\cite{hoover2015comparison} to avoid discontinuous forces (as in hard-wall reflections) and to maintain smooth phase-space dynamics.

Figure 1 shows a contour plot of the potential $\phi(r_2)$ with the scatterers located at $(\pm 1, \pm 1)$. The plot illustrates how the potential rises steeply near the corner, ensuring smooth confinement. After calculating the forces at each step, the particle will be placed back into the cell. When the particle crosses the cell boundaries, it is repositioned within the $2\times2$ unit cell as follows:
\begin{align}
x = x + 2, && \text{if } x < -1, \\
x = x - 2, && \text{if } x > +1, \\
y = y + 2, && \text{if } y < -1, \\
y = y - 2, && \text{if } y > +1.
\end{align}

We adopt the same initial values as in the original work of the cell model $ \{ x, y, p_x, p_y, \zeta_x, \zeta_y \} = \{0.0, 0.0, 0.6, 0.8, 0.0, 0.0\}$ ~\cite{hoover2015comparison}, which yields an initial total energy of 0.5, as the particle begins in the field-free region, $r_w = (0,0)$. In the absence of thermostats (Hamiltonian case), the system can exhibit both chaotic and non-chaotic trajectories, depending on initial conditions. Our simulations focus on chaotic trajectories to later examine the effect of thermostats on these dynamics.

\begin{figure}[t]
    \centering
    \includegraphics[width=\linewidth]{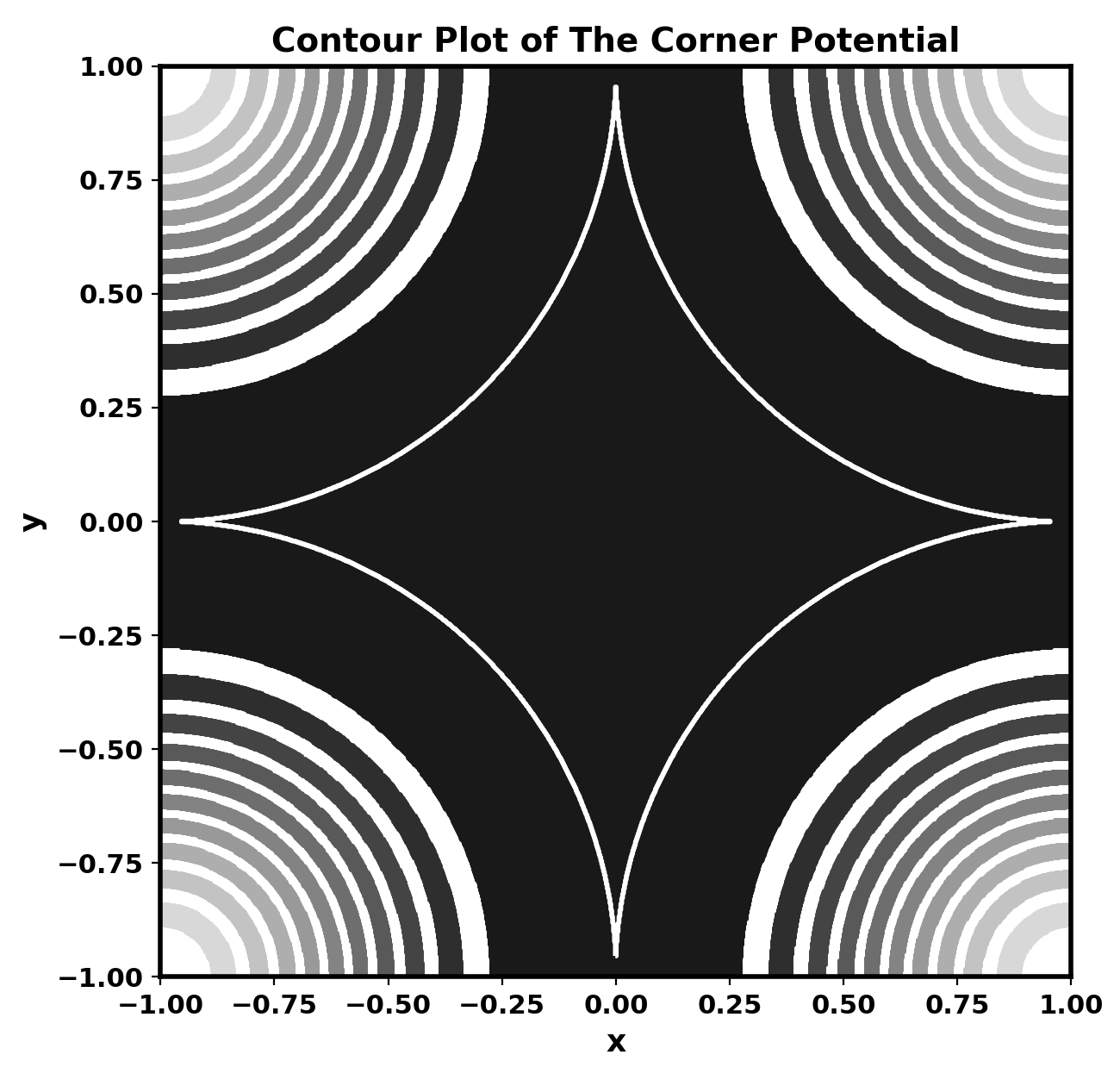}
    \caption{Contour plot of the corner potential $\phi(r)$ within a square cell of side length~2 with repulsive corners at $(\pm 1, \pm 1)$. The potential is calculated as the sum of four truncated radial functions centered at the corners, each contributing $\sum_{i=1}^{4}  ~[1 - |r_w - r_i|^2]^4 ~ \text{for} ~ |r_w - r_i| < 1$. Contours are plotted using ten shaded bands in $\phi$ of width~0.05: [0.00--0.05], [0.10--0.15], [0.20--0.25], \ldots, [0.90--0.95], leaving white space in between bands to highlight curvature. The $\phi=0$ contour emerges where the potentials from adjacent corners intersect.}
    \label{fig:corner_potential}
\end{figure}

\subsubsection{Hamiltonian Reference Case}
we also consider a strictly Hamiltonian version with no thermostat terms. In this case $\zeta_x = \zeta_y = 0$, and
\begin{align}
\dot{x} &= p_x, \quad \dot{p}_x = -\frac{\partial \phi}{\partial x}, \\
\dot{y} &= p_y,  \quad \dot{p}_y = -\frac{\partial \phi}{\partial y}.
\end{align}
The total Hamiltonian energy $E = \frac{1}{2}\left( p_x^2 + p_y^2 \right) + \phi(r_w),$ where $\phi(r_w) = \phi(x,y)$, is conserved throughout the dynamics.

\subsection{Numerical Methods}

\subsubsection{Integrator}
We use fourth‐order Runge–Kutta (RK4) to integrate the four Hamiltonian and the six nonequilibrium equations of motion. A fixed timestep $\Delta t$ is chosen (typically $\Delta t$ = 0.001), verified for convergence by checking final results over smaller $\Delta t$. Depending on the run, we integrate as many as $10^{11}$ steps to ensure the system reaches a steady state and the Lyapunov exponents converge within an uncertainty of 0.001. Data are stored every 1000 steps for analysis, corresponding to integer-valued times.

\subsubsection{Lyapunov Exponent Calculations}
Lyapunov exponents $\{\lambda_i\}$ quantify the average exponential rates at which nearby trajectories in phase space diverge (or converge) over time. For a D-dimensional dynamical system, there are D exponents, each describing the growth rate along a principal axis of perturbation. A positive exponent indicates exponential $(\propto e^{\lambda t})$ sensitivity to initial conditions, characteristic of chaos, while negative exponents reflect contraction, and zero exponents correspond to neutral (conserved) directions. Formally, if $\delta (t)$ is an infinitesimal perturbation to the system’s state, its growth is governed by:

\begin{align}
\lambda = \lim_{t\to\infty} \frac{1}{t} ln\frac{|\delta(t)|}{|\delta(0)|}
\end{align}

where $| .|$ is the length of the perturbation. Following the approach of Benettin’s group \cite{benettin1976kolmogorov}, we compute the full spectrum by evolving a set of orthonormal perturbation vectors (satellite trajectories) along the reference trajectory. At each integration step, the Gram-Schmidt orthonormalization is applied to measure the divergence/convergence of tangent vectors. The time‐averaged logarithmic growth rates converge to the Lyapunov exponents  $\{\lambda_i\}$  with an uncertainty of 0.001. 

\section{Results and Discussion}
We first show the trajectory of the Hamiltonian cell-model oscillator (without thermostats), integrated for $10^6$, steps in figure 2. Next, we show the trajectories of cell model under the two-temperature Nos\'e–Hoover thermostat. We keep $T_x$ fixed at 0.5 and decrease $T_y$ from 0.5 to 0.05 in ten equal steps of 0.05 unit. Here we show three representative trajectories of $(T_x,T_y)$: (i) near‐equilibrium (0.50, 0.45), (ii) moderate difference (0.50, 0.25), and (iii) large difference (0.50, 0.05) in figure 3, respectively. In all nonequilibrium cases of varying $T_x$   and $T_y$ , the particle gains higher energy due to the thermostat coupling to its motion and/or kinetic energy, allowing it to overcome the repulsive potential barriers and reach the corners. In addition, one can observe that the particle tends to move along x-axis due to $T_x>T_y$ . This is more evident in $T_x$ = 0.50, $T_y$ = 0.05, where the particle predominantly moves in the x direction.

\begin{figure}[t]
\centering
\includegraphics[width=0.32\linewidth]{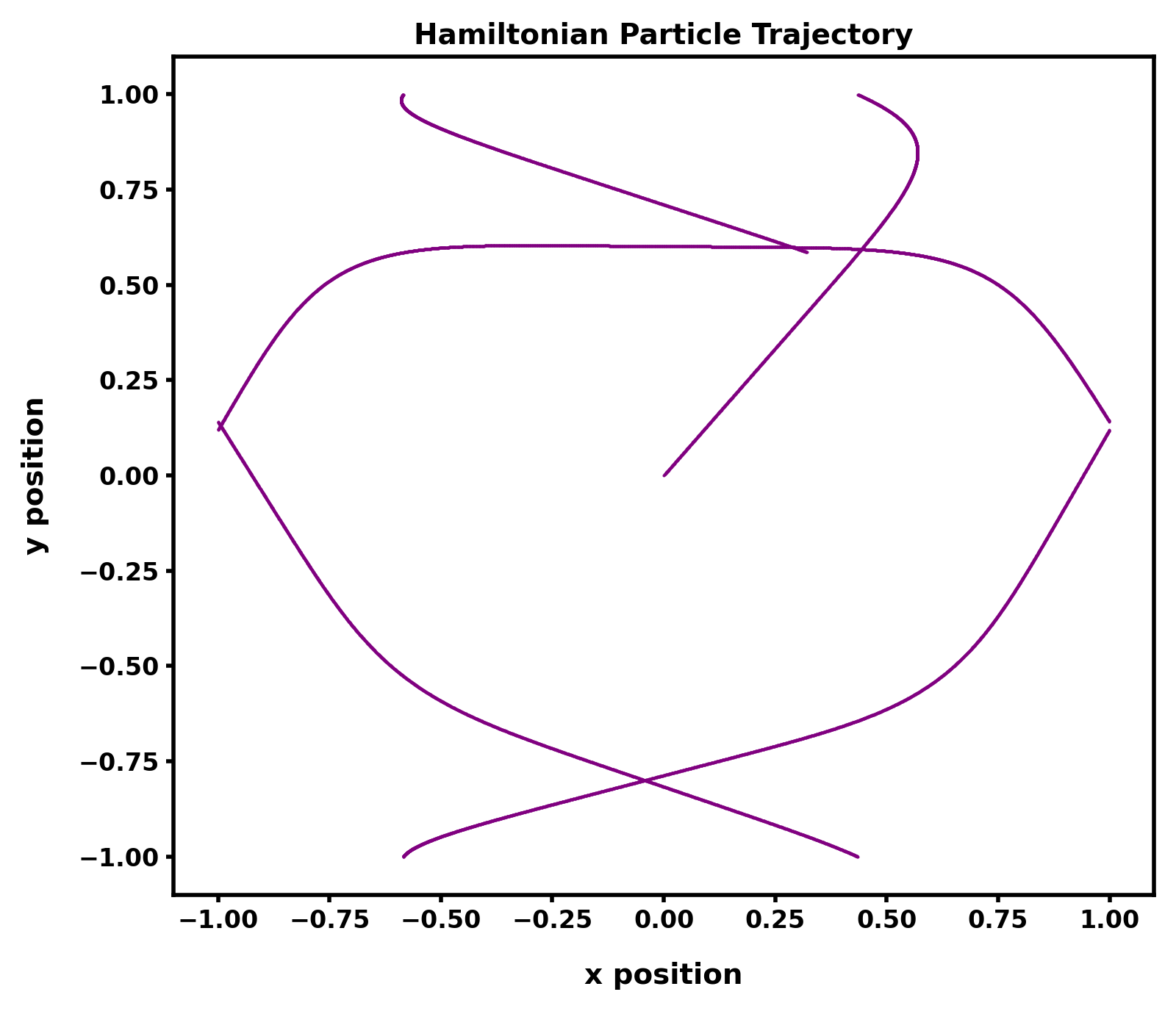}
\includegraphics[width=0.32\linewidth]{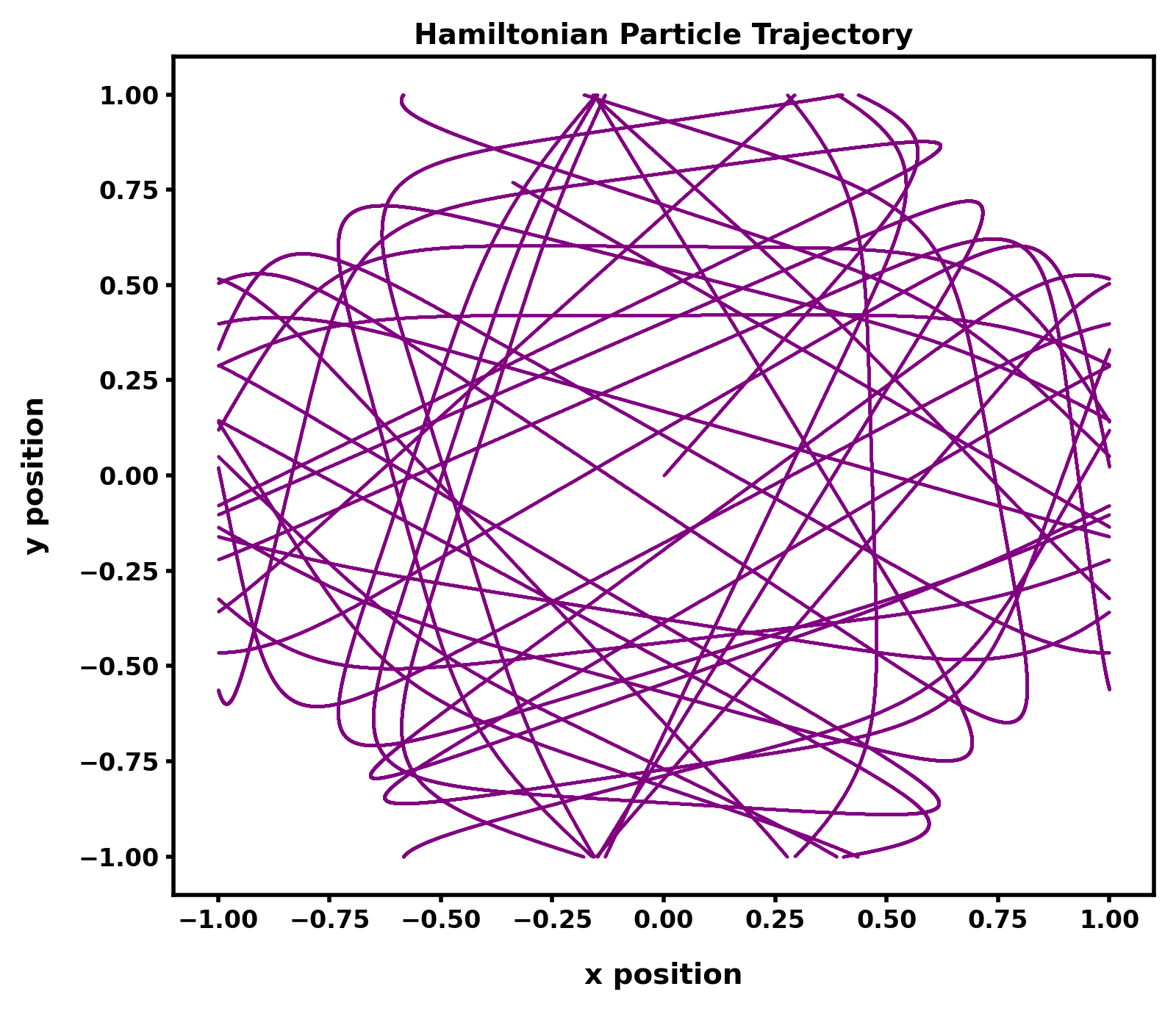}
\includegraphics[width=0.32\linewidth]{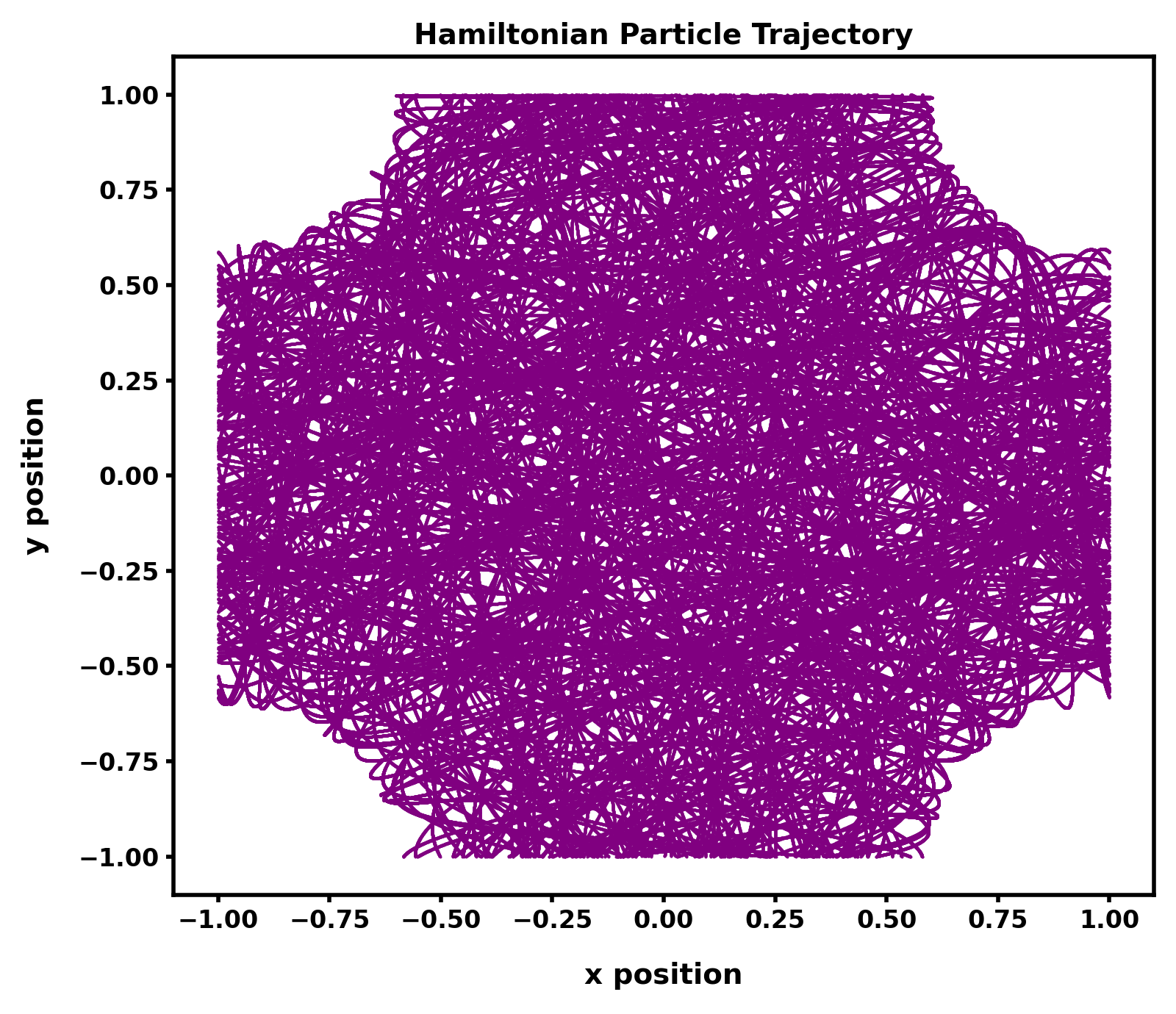}
\caption{Particle trajectories under Hamiltonian mechanics, integrated for $10^4$ (left), $10^5$ (middle), and $10^6$ (right) steps.
}
\label{fig:hamiltonian_traj}
\end{figure}

\begin{figure}[t]
\centering
\includegraphics[width=0.32\linewidth]{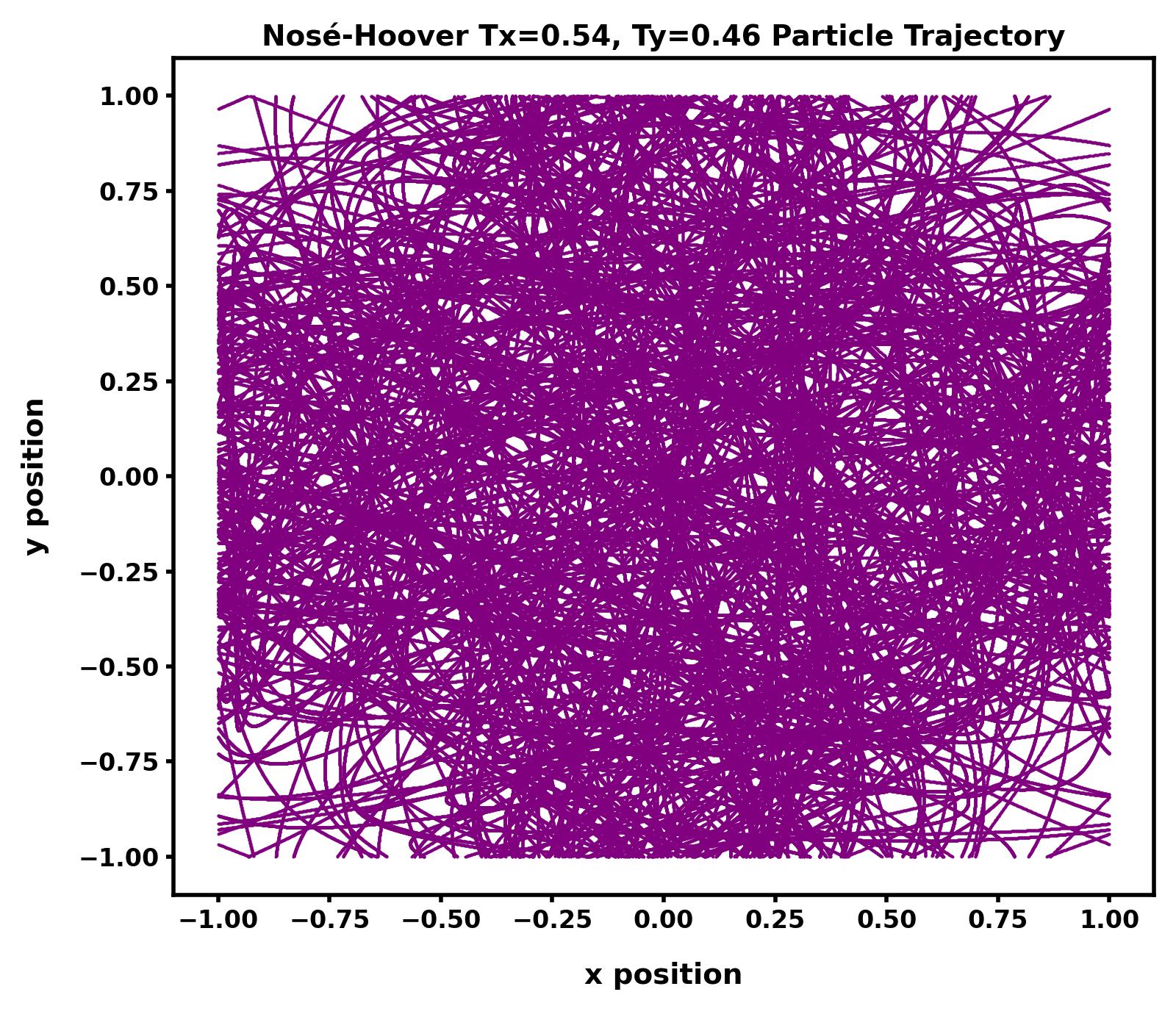}
\includegraphics[width=0.32\linewidth]{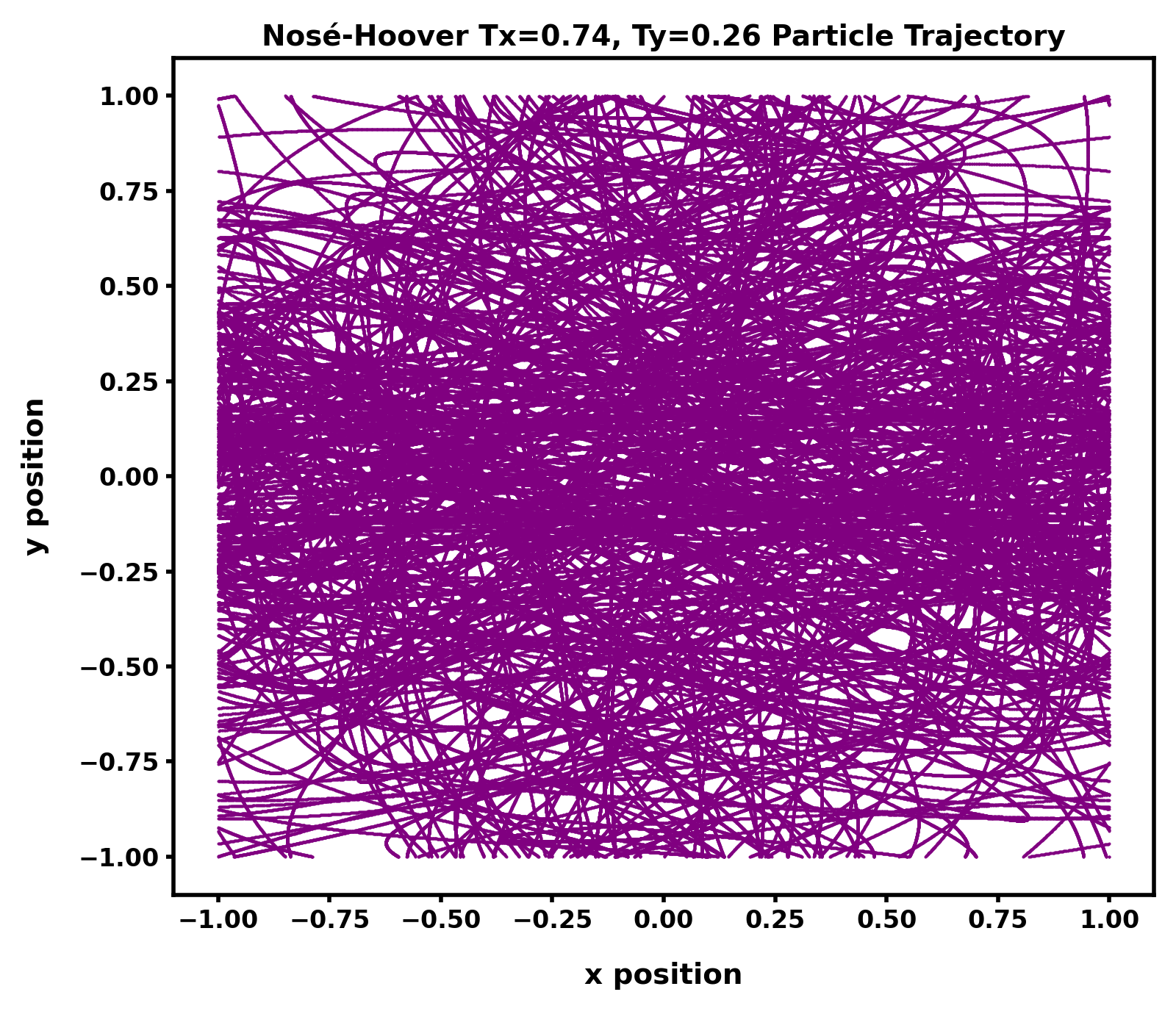}
\includegraphics[width=0.32\linewidth]{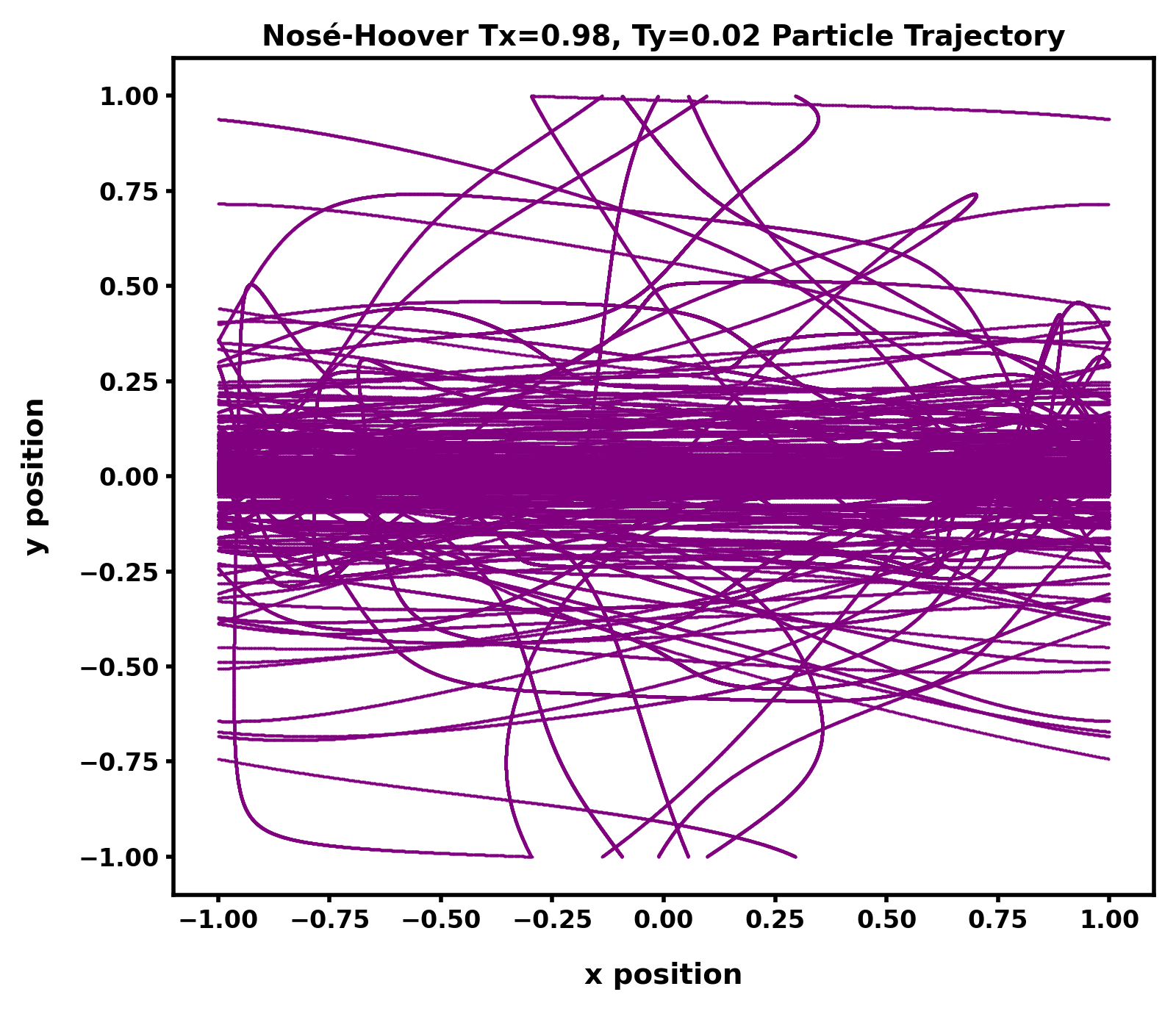}
\caption{Particle trajectories under two-thermostat Nos\'e--Hoover mechanics integrated for $10^6$ steps with $(T_x,T_y)$: left) near-equilibrium $(0.50, 0.45)$, middle) moderate difference $(0.50, 0.25)$, and right) large difference $(0.50, 0.05)$.
}
\label{fig:NH_traj}
\end{figure}

\subsection{Chaotic Dynamics and Lyapunov Exponents}
To illustrate that the system exhibits chaotic dynamics in a nonequilibrium steady state (NESS), we follow each Lyapunov exponent as a function of simulation time. Each run was extended to $10^{11}$ steps, reducing the one-$\sigma$ uncertainty of each $\lambda_i$ to less than 0.001. Supplementary Figure S1 shows the Lyapunov spectrum of all nonequilibrium cases. The Lyapunov exponents for all $T_x,T_y$ sets are shown in Supplementary table S1. Each spectrum contains one to three positive exponents, confirming chaos. We also note that $\lambda_3$ and $\lambda_4$, which are close to zero within one-$\sigma$ uncertainty, asymptotically vanish in the long-time limit, reflecting the marginally stable directions of the dynamics. Next, $\Lambda$ denotes the total phase-space contraction rate. Here it is defined as the sum of six Lyapunov exponents \cite{posch1989equilibrium}, $\Lambda = \sum_{i=1}^6 \lambda_i$ . In Nos\'e–Hoover dynamics, $\Lambda$ quantifies the net dissipation imposed by the thermostats and characterizes how the volume of a phase-space element changes over time under their influence. Table 1 summarizes the values of $\Lambda$ calculated from the final converged exponents $\{\lambda_1, \ldots, \lambda6\}$. The contraction rate, $\Lambda$, becomes increasingly negative as $|T_x-T_y|$ grows, e.g. -0.001 for the near-equilibrium case (0.50, 0.45) to -0.139 for the strongly driven pair (0.50, 0.05), indicating stronger dissipation farther from equilibrium. We calculated the Kaplan–Yorke dimension, $D_{KY} = j + \frac{\sum_{i=1}^{j} \lambda_i}{|\lambda_{j+1}|}$, where $j$ is the largest index for which $\sum_{i=1}^j\lambda_i \geq 0$ . We find $5.0 < D_{KY} < 6.0$, indicating that although the full phase space is six-dimensional, the driven system’s attractor occupies a fractal set of dimension < 6. This monotonic decrease indicates that stronger thermostat anisotropy compresses the attractor further below the six‑dimensional embedding space, in agreement with intuition that larger heat‑flux driving yields greater dissipation and lower dimensionality. Alternative but equivalent formulations of the contraction rate, expressed directly in terms of thermostat variables, are provided in Appendix.

\begin{table}[t]
\caption{Phase-space contraction rate and Kaplan--Yorke dimension, $D_{\rm KY}$, for each thermostat pair in the anisotropic scan. The increasing magnitude of $\Lambda$ and decreasing $D_{\rm KY}$ reflect greater dissipation and thinner attractors as $T_y$ decreases. Values are based on simulations with $10^{11}$ integration steps. In the thermostatted cases, $T_x = 0.5$.
}
\label{tab:KY_dimension}
\centering
\begin{tabular}{cccc}
\hline\hline
$T_y$ & $\delta = 0.5 - T_y$ & $\Lambda$ & KY dimension \\
\hline
0.50 & 0.00 &  0.000  & 6.00 \\
0.45 & 0.05 & -0.001  & 5.99 \\
0.40 & 0.10 & -0.004  & 5.99 \\
0.35 & 0.15 & -0.009  & 5.98 \\
0.30 & 0.20 & -0.019  & 5.95 \\
0.25 & 0.25 & -0.034  & 5.91 \\
0.20 & 0.30 & -0.052  & 5.86 \\
0.15 & 0.35 & -0.072  & 5.79 \\
0.10 & 0.40 & -0.096  & 5.69 \\
0.05 & 0.45 & -0.139  & 5.49 \\
\hline
Hamiltonian & 0.00 & 0.000 & 6.00 \\
\hline\hline
\end{tabular}
\end{table}

\subsection{Poincar\'e Sections Analysis}

\begin{figure}[!t]
  \centering
  \includegraphics[height=0.22\textheight,keepaspectratio]{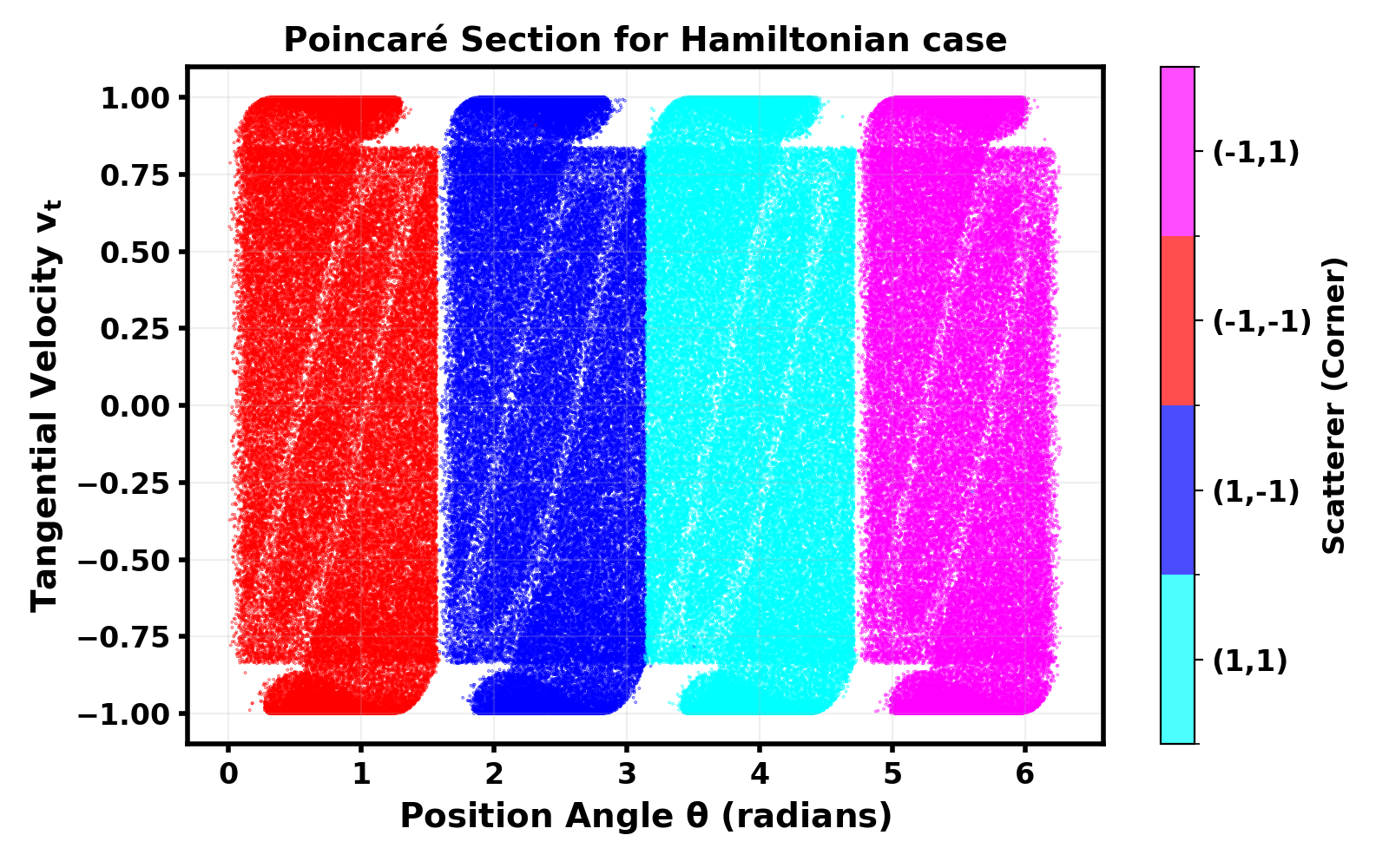}\par\smallskip
  \includegraphics[height=0.22\textheight,keepaspectratio]{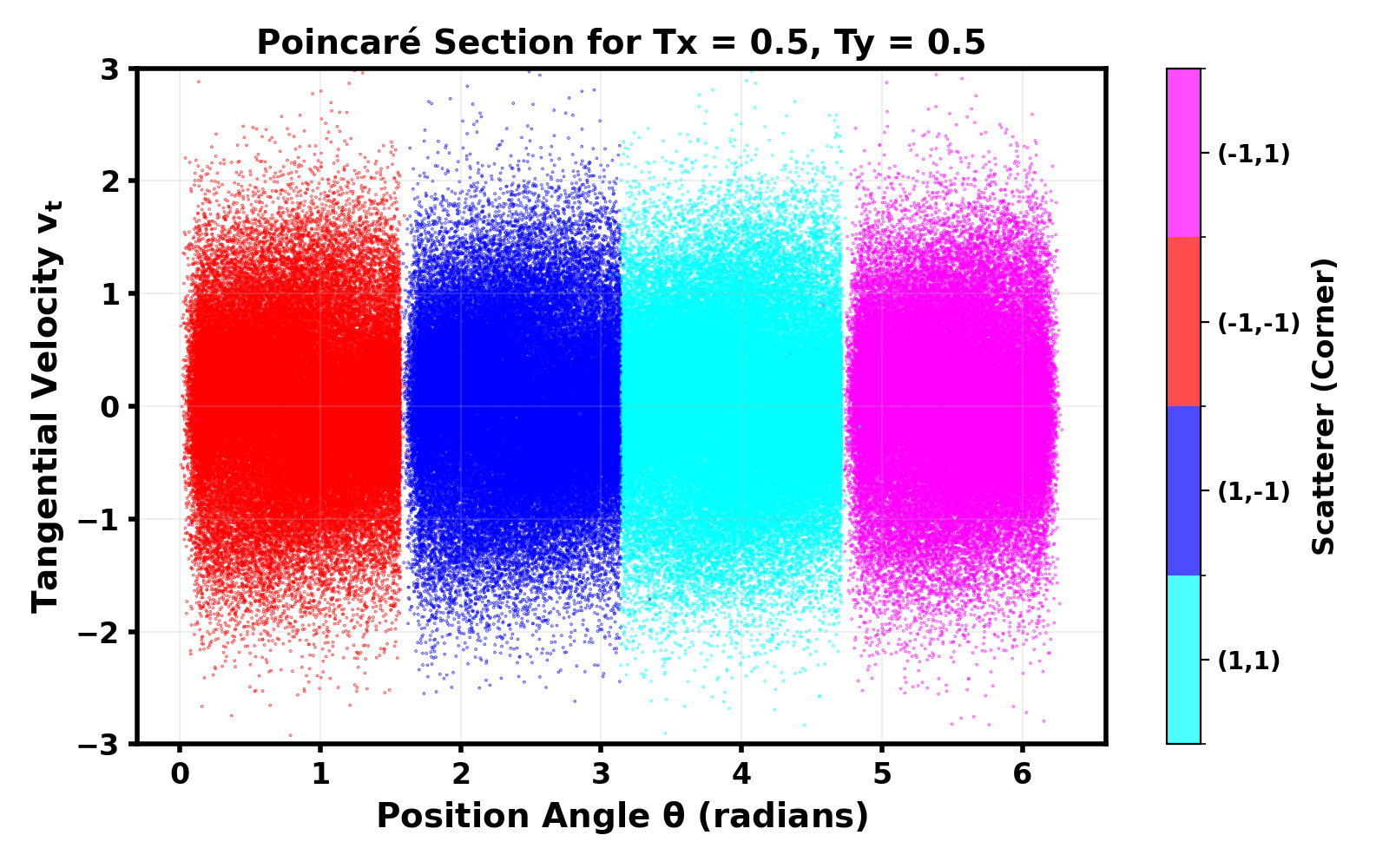}\par\smallskip
  \includegraphics[height=0.22\textheight,keepaspectratio]{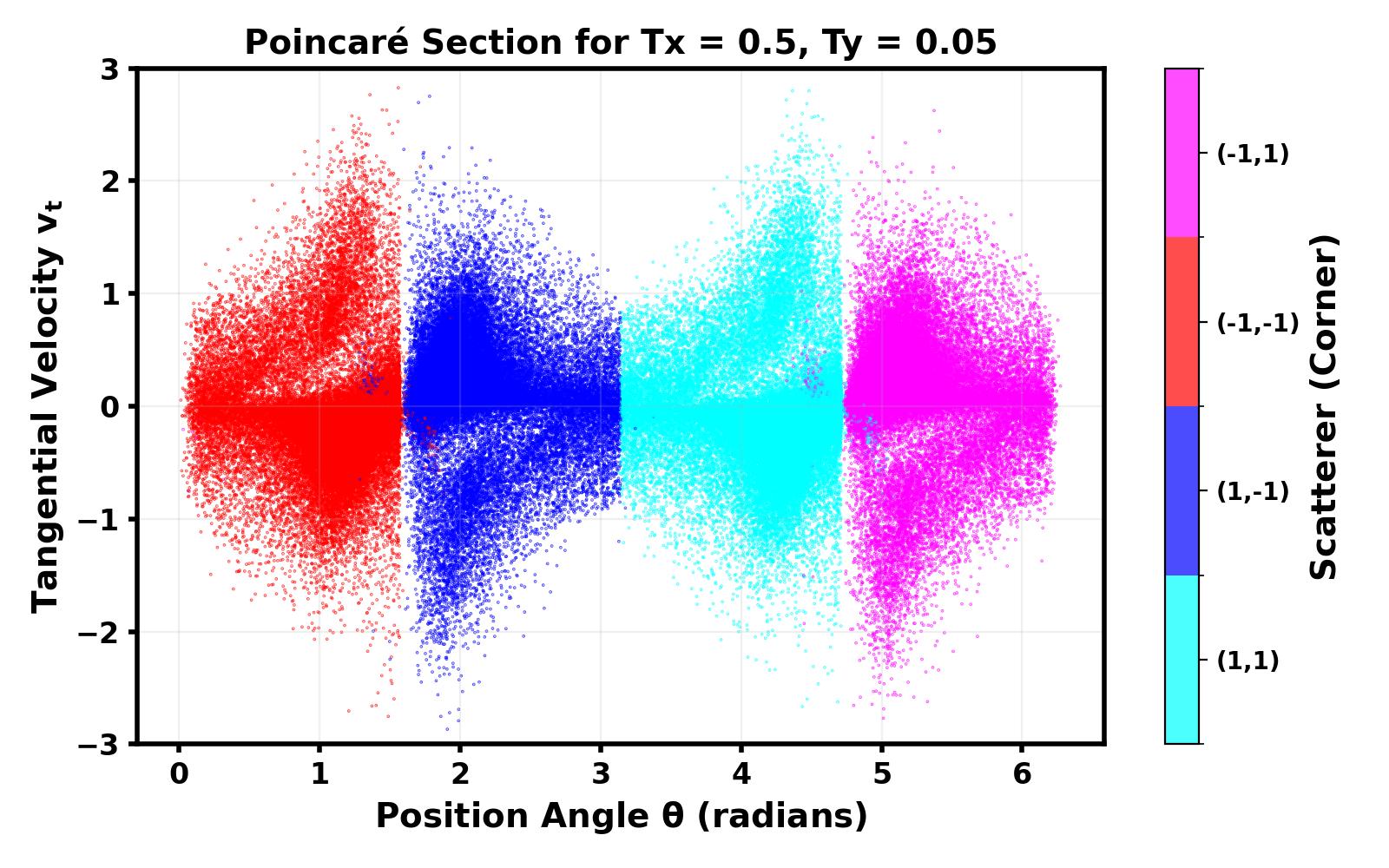}
  \caption{Poincar\'e sections showing the position angle versus tangential velocity at the boundary of the zero-potential region for the Hamiltonian, equilibrium $(T_x=T_y=0.5)$, and driven $(T_x=0.5,\,T_y=0.05)$ cases. Each panel demonstrates the crossings at the four scatterers, illustrating the symmetry of the system under different thermal conditions. The Hamiltonian and equilibrium cases exhibit fourfold symmetry consistent with the system’s geometry and ergodic exploration of phase space, while the driven case shows a pairwise $180^\circ$ rotational symmetry, indicating shear deformation under anisotropic drive. Each simulation was run for $10^{11}$ steps with the same initial conditions described in Sec.~\ref{sec:model}, ensuring consistency with all other simulations in this study.}
  \label{fig:poincare}
\end{figure}

To complement our Lyapunov exponent and Kaplan--Yorke dimension analysis, we constructed Poincar\'e sections by recording the position angle and tangential velocity of the particle each time it crossed the boundary of the zero-potential region~\cite{balint2023periodic}, as shown in Fig.~\ref{fig:poincare}. The position angle was defined relative to the center of each scatterer, while the tangential velocity was computed as the component of the particle's velocity along the local tangent at the crossing point. 

In the Hamiltonian and equilibrium ($T_x = T_y = 0.5$) cases, the Poincar\'e sections show a fourfold symmetry consistent with the system’s geometry, indicating ergodicity alongside the chaotic dynamics verified by the Lyapunov exponents. In contrast, under driven conditions ($T_x > T_y$), the Poincar\'e sections show that diagonally opposite corners exhibit nearly identical distributions, consistent with an approximate $180^\circ$ rotational symmetry of the system under anisotropic driving. In other words, if we rotate the entire system by $180^\circ$ around the box center, the entire Poincar\'e section pattern remains approximately invariant, showing rotational symmetry across the center, while each diagonal pair itself is nearly identical. Although the anisotropic thermostatting breaks full equilibrium symmetry, the system retains this reduced rotational symmetry even far from equilibrium, reflecting the constraints imposed by the geometry and boundary conditions. While visually informative, the Poincar\'e section primarily serves as a qualitative check, and we rely on the Lyapunov exponents and fractal dimensionality for quantitative characterization and reproducibility of chaos in the system.

\subsection{Entropy Production and Dissipation}

In deterministic, time-reversible thermostatted dynamics, the negative of the total Lyapunov sum often coincides with the entropy production rate in the heat baths,
\begin{equation}
\dot{S}_{\mathrm{prod}} = -k \Lambda,
\end{equation}
where $k$ is Boltzmann’s constant~\cite{posch1989equilibrium}. Thus, the more negative $\Lambda$ becomes, the larger the irreversible heat current from the system to the thermostats, and the higher the dissipation. Table~\ref{tab:KY_dimension} illustrates this trend: near-equilibrium runs ($T_x \approx T_y$) show small $|\Lambda|$ and mild entropy production, while large $\Delta T$ cases yield significantly higher $\dot{S}_{\mathrm{prod}}$. When $T_x \neq T_y$, the thermostats drive net heat flow from the hotter degree of freedom to the cooler one, causing phase-space contraction onto a nonequilibrium attractor. For example, at $(T_x, T_y) = (0.50, 0.05)$ we observe a strongly negative total exponent, suggesting heat flow from the $x$ (hot) degree of freedom to $y$ (cold), which leads to greater dissipation. To summarize the effect of thermal anisotropy on dissipation, we fit the total phase-space contraction rate $\Lambda$ to two different models:
\begin{enumerate}
\item A power-law model of the form $\Lambda = A \delta^{B}$, 
\item A quadratic-plus-quartic model $\Lambda = A \delta^2 + B \delta^4$,
\end{enumerate}
where $\delta = 0.5 - T_y$ and $T_x = 0.5$ is held constant. A nonlinear least-squares fit to the power-law form yields the empirical relation:
\begin{equation}
\Lambda = -0.95 \, \delta^{2.44}.
\end{equation}

According to linear response theory, entropy production in near-equilibrium systems is given by
\begin{equation}
\dot{S}_{\mathrm{prod}} = \sum_i J_i X_i,
\end{equation}
where $J_i$ and $X_i$ are thermodynamic fluxes and forces, respectively. In the case of heat flow, the flux $J$ (e.g., heat current) is linearly related to the driving force $X$ (e.g., temperature anisotropy $\delta = T_x - T_y$) via a transport coefficient $L$, such that $J = L \delta$. This leads to a quadratic dependence of entropy production,
\begin{equation}
\dot{S}_{\mathrm{prod}} = L \delta^2,
\end{equation}
and consequently $\Lambda \propto -\delta^2$, since phase-space contraction is a measure of dissipation~\cite{posch1989equilibrium,evans2008statistical}. However, our simulations show a steeper scaling in the form of $\delta^{2.44}$.

\begin{figure}[t]
    \centering
    \includegraphics[width=\linewidth]{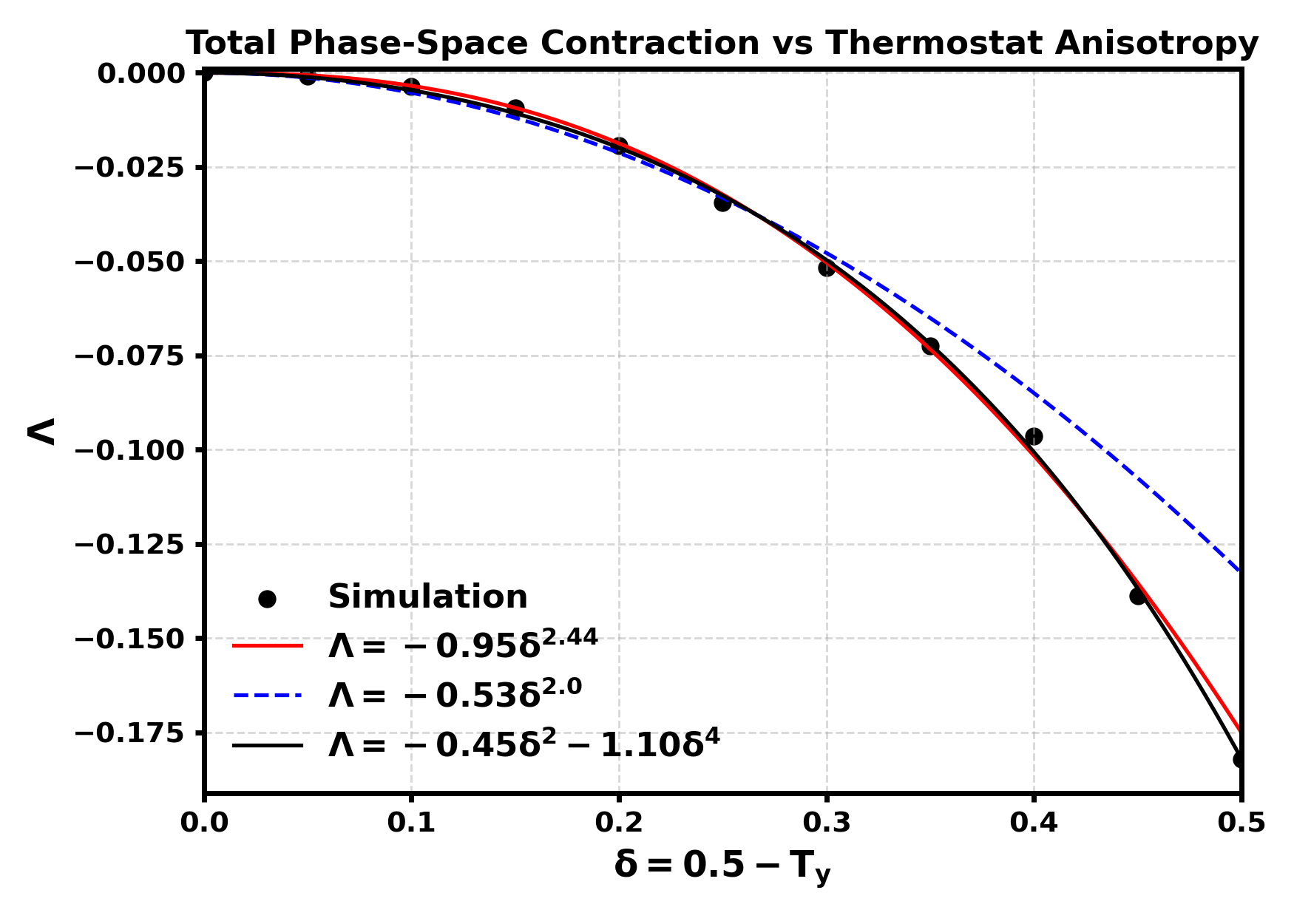}
    \caption{%
    Total phase-space contraction rate, $\Lambda$, as a function of thermostat anisotropy $\delta = 0.5 - T_y$ with fixed $T_x = 0.5$. Simulation data with error bars are shown in black. Three fits are overlaid for comparison: a quadratic model, $\Lambda = -0.51 \, \delta^2$ (blue), consistent with linear-response theory; a power-law fit, $\Lambda = -0.95 \, \delta^{2.44}$ (red), suggesting nonlinear scaling; and a combined quadratic~+~quartic, $\Lambda = -0.45 \, \delta^2 - 1.10 \, \delta^4$ (black), offering an alternative nonlinear description. While both nonlinear fits capture the simulation data well across the full $\delta$ range, the power-law is slightly better at low $\delta$, and the quartic term improves agreement at high $\delta$.
    }
    \label{fig:Lambda_fits}
\end{figure}

A quartic extension to the quadratic form, motivated by a Taylor expansion of $\Lambda$ in $\delta$, that is consistent with linear-response theory near equilibrium:
\begin{equation}
\Lambda = -0.45 \, \delta^2 - 1.10 \, \delta^4,
\end{equation}
fits the data comparably well, particularly at large $\delta$. While the power-law form provides a slightly better fit at small $\delta$, the quartic model better captures behavior in the strongly driven regime. We therefore present both, leaving the precise functional form open to interpretation (see Fig.~\ref{fig:Lambda_fits}).

This deviation from strict quadratic scaling highlights the limitations of linear-response theory in strongly nonequilibrium regimes. Whether this nonlinearity reflects underlying multifractal phase-space structure or emergent transport phenomena remains an open question.

\begin{figure}[t]
    \centering
    \includegraphics[width=\linewidth]{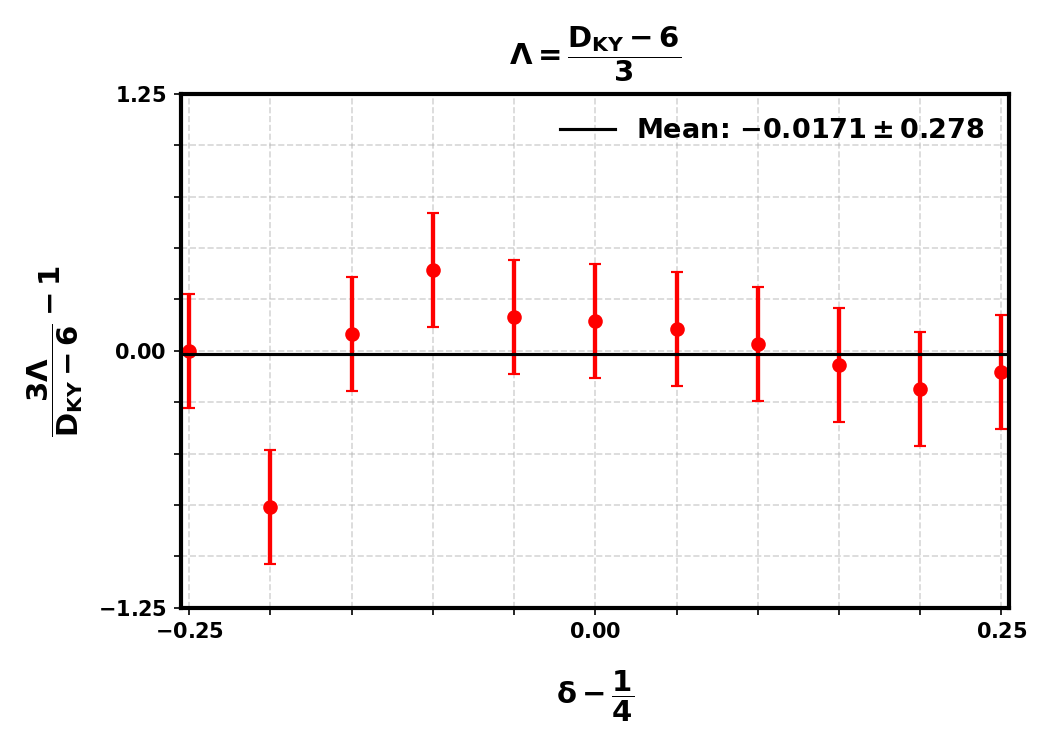}
    \caption{%
    Residual deviation from the phase space contraction--dimensionality relation, plotted as a function of centered thermostat anisotropy, $\delta - 1/4$. The scaled residuals are defined as $3\Lambda/(D_{\rm KY} - 6) - 1$. The mean residual (solid line) is $-0.017$ with a standard deviation of $\pm 0.278$. This normalization points to a hypothesis that dissipation in the $y$-direction, governed by $y$, $p_y$, and $\zeta_y$, scales linearly with the corresponding phase-space dimensionality loss. In order to obtain the $T_y = 0$ limit, the final point at $\delta = 0.5$ is obtained from linear extrapolation of both $\Lambda$ and $D_{\rm KY}$ as functions of $\delta$.
    }
    \label{fig:Lambda_DKY}
\end{figure}

We also investigate the relationship between dissipation and dimensionality loss $D_{\rm KY} - 6$ by testing the empirical scaling relation:
\begin{equation}
\Lambda \approx \frac{D_{\rm KY} - 6}{3}.
\end{equation}
The factor of 3 arises from anisotropic dissipation that is confined to the $y$-direction, which involves three phase-space variables: $y$, $p_y$, and $\zeta_y$. To evaluate this relation, we plotted the normalized deviation $\frac{3\Lambda}{D_{\rm KY} - 6} - 1$ as a function of $\delta - \frac{1}{4}$, which centers the temperature anisotropy range at the origin. As shown in Fig.~\ref{fig:Lambda_DKY}, most of the residuals remain within one standard deviation, providing support for the linear scaling between dissipation and dimensionality loss. The first point at $\delta = 0$ corresponds to the equilibrium (Hamiltonian) limit, while the final point at $\delta = 0.5$ is obtained from linear extrapolation.

It is worth mentioning that the attempt to use Nos\'e--Hoover thermostatting at very low temperatures provokes the `Norse troll under the bridge', namely, the fearsome Toda demon (see Ref.~\cite{holian1995thermostatted}). Nos\'e--Hoover thermostatting is a Black Hole singularity at $T_y = 0$, as can be seen from the equation of motion of the thermostatting variable in the $y$-dimension. As a result, the $T_y = 0$ limit of both the $\Lambda$ and $D_{\rm KY}$ data must be obtained via linear extrapolation as functions of $\delta$.

\subsection{Thermostat statistics and non-Gaussian moments}

For a strictly canonical (Gaussian) one-dimensional momentum distribution, the normalized fourth and sixth moment ratios are:
\begin{equation}
R_4 = \frac{\langle p^4 \rangle}{\langle p^2 \rangle^2} = 3, 
\qquad
R_6 = \frac{\langle p^6 \rangle}{\langle p^2 \rangle^3} = 15.
\end{equation}

Deviations from these canonical values measure how far the steady-state single-particle dynamics is driven away from equilibrium. We computed $R_4$ and $R_6$ separately for $p_x$ and $p_y$ to quantify the anisotropic impact of the thermostats. To estimate one-$\sigma$ uncertainties in $R_4$ and $R_6$, the full trajectory was divided into 20,000-step blocks, and the standard deviation of each moment ratio was computed among blocks. For the near-equilibrium case $(T_x = 0.50,\, T_y = 0.45)$, both $R_4^y$ and $R_6^y$ remain within a few percent of the canonical values, reflecting quasi-Gaussian dynamics. As the thermal anisotropy increases $(T_y \to 0.25)$, these ratios rise. Under strong driving $(T_y = 0.05)$, the sixth moment $R_6^y$ increases to $\sim$160, reflecting rare but extreme fluctuations in $p_y$. While most moment ratios remain statistically indistinguishable from Gaussian values within uncertainty, only the final few thermostat pairs show significant deviations. The complete set of values and uncertainties is provided in Supplementary Table~S2. The momentum tails complement the growing entropy production, providing statistical evidence of the system’s deviation from equilibrium.

\subsection{Time Reversibility}

Although our two-thermostat system is both dissipative $\left( \sum_{i=1}^j \lambda_i < 0 \right)$ and nonequilibrium ($T_x \neq T_y$), it remains time-reversible in the sense that changing the signs of $(p_x, p_y)$ and $(\zeta_x, \zeta_y)$ along with reversing time $t \rightarrow -t$ leaves the equations of motion invariant. More specifically, because the friction-like term $-\zeta_x p_x$ changes sign twice ($-1 \cdot -1 = +1$) and since $\zeta_x$ depends only on $p_x^2$, the flow is symmetric under this transformation. As a result, if one were to record the phase-space trajectory forward in time and then instantaneously reverse momenta and thermostats, evolving with negative time, the system would retrace its path in phase space. This time-reversible but dissipative property underscores how microscopic time reversibility can coexist with a fractal, contracting attractor at the macroscopic level.

\section{Conclusions}

We revisited the two-temperature Nos\'e--Hoover cell model embedded in a cell with four repulsive corners to obtain a simple deterministic model for exploring anisotropic thermostatting. Owing to the model’s low dimensionality, we were able to compute the full six-dimensional Lyapunov spectrum with high numerical precision (uncertainty $< 0.001$), confirming that the system is chaotic for all thermostat pairs studied. The total phase-space contraction rate $\Lambda$ becomes increasingly negative as the imposed temperature difference grows, in agreement with the entropy-production formula $\dot{S}_{\mathrm{prod}} = -k\Lambda$. The approximate Kaplan--Yorke estimate of the attractor dimension decreases from the equilibrium value $D = 6$ to $D \approx 5.5$ at $(T_x, T_y) = (0.50, 0.05)$, demonstrating that stronger driving compresses the accessible phase space. The dissipation scales with thermostat anisotropy in a nonlinear fashion. Both a power-law fit $\Lambda \propto -\delta^{2.44}$ and a quadratic-plus-quartic form provide good descriptions of the data, with the latter yielding better agreement in the strongly driven regime and consistency with linear-response theory near equilibrium. We also find an empirical relation, $\Lambda \approx (D_{\rm KY} - 6)/3$, showing the role of three $y$-directed phase-space variables in anisotropic dissipation. Complementary analyses of momentum distributions through the fourth and sixth moments ($R_4$, $R_6$) further support this interpretation. Because the motion equations remain strictly time-reversible, the model provides a pedagogical example of the way microscopic reversibility coexists with macroscopic dissipation. Future work could test the two-temperature thermostat in other cases, perhaps in three dimensions, and explore whether analogous fractal contraction appears in many-particle versions of the two-temperature thermostat.

\appendix*
\section{Phase–space contraction and entropy production}
\vskip -0.2cm
We present two complementary derivations of the phase-space contraction rate $\Lambda$: one based on thermostat equations of motion and conservation of energy, and one based directly on the flow compressibility (divergence of the phase-space velocity field). Both routes lead to the same steady–state identity (with $k=1$). We emphasize that these identities are used here solely as a validation of the Lyapunov–based evaluation of $\Lambda$ employed in the main text.\\

\paragraph{mechanical route\\\\}

The total energy, $E$, for the two-temperature Nos\'e-Hoover thermostat in 2D (with $m=1$):
\begin{equation}
E(x,y,p_x,p_y) = \frac{1}{2}\big(p_x^2 + p_y^2\big) + U(x,y).
\end{equation}
by the chain rule:
\begin{align}
\dot E
&= \frac{\partial E}{\partial x}\dot x
 + \frac{\partial E}{\partial y}\dot y
 + \frac{\partial E}{\partial p_x}\dot p_x
 + \frac{\partial E}{\partial p_y}\dot p_y \\
&= \underbrace{(\partial_x U)\,\dot x}_{(A)}
 + \underbrace{(\partial_y U)\,\dot y}_{(B)}
 + \underbrace{p_x\,\dot p_x}_{(C)}
 + \underbrace{p_y\,\dot p_y}_{(D)}. \label{eq:Edot-expand}
\end{align}
\vskip -0.1cm
Insert \eqref{eq:eom1}--\eqref{eq:eom4}:
\begin{align}
(A)+(B) &= (\partial_x U)\,p_x + (\partial_y U)\,p_y,\\
(C)+(D) &= p_x(F_x - \zeta_x p_x) + p_y(F_y - \zeta_y p_y).
\end{align}
since \(F_x=-\partial_x U\) and \(F_y=-\partial_y U\), the conservative contributions cancel, hence
\begin{equation}
\boxed{\;\dot E = -\,\zeta_x\,p_x^2 \;-\; \zeta_y\,p_y^2 = \dot Q_x + \dot Q_y. \;}
\label{eq:Edot-final}
\end{equation}
We consider $ \dot Q_x + \dot Q_y$ as heat flux into system. Entropy production rate of baths:
\begin{equation}
\dot S_{\mathrm{prod}} = \frac{\dot Q_x}{T_x} + \frac{\dot Q_y}{T_y} = \frac{\zeta_x p_x^2}{T_x} + \frac{\zeta_y p_y^2}{T_y}. \label{eq:Sprod-start}
\end{equation}
Using equations \eqref{eq:eom5} - \eqref{eq:eom6} (with $m = k= \tau_x = \tau_y = 1$), $\frac{p_i^2}{T_i} = 1 + \dot\zeta_i$, where  \(i\in\{x,y\}\):
\begin{align}
\frac{\zeta_i p_i^2}{T_i} = \zeta_i\,(1+\dot\zeta_i)
= \zeta_i + \frac{1}{2}\,\frac{d}{dt}\big(\zeta_i^2\big),
\end{align}
\begin{align}
\dot S_{\mathrm{prod}} = \zeta_x + \zeta_y + \frac{d}{dt}\left[\frac{1}{2}\big(\zeta_x^2 + \zeta_y^2\big)\right]. \label{eq:Sprod-split}
\end{align}
In NESS, a stationary process, $f(t) = \tfrac{1}{2}\big(\zeta_x(t)^2+\zeta_y(t)^2\big)$ has a time-independent distribution and is bounded in mean, hence its long–time average derivative vanishes:
\begin{equation}
\begin{aligned}
\left\langle\frac{d}{dt} f(t) \right\rangle 
&= \lim_{T\to\infty}\frac{1}{T}\!\int_0^T \frac{d}{dt} f(t)\,dt \\
&= \lim_{T\to\infty}\frac{f(T)-f(0)}{T} = 0
\end{aligned}
\label{eq:total-deriv-vanish}
\end{equation}
Taking the long-time average of \eqref{eq:Sprod-split}:
\begin{equation}
\boxed{\;
\left\langle \dot S_{\mathrm{prod}} \right\rangle = \left\langle \zeta_x + \zeta_y \right\rangle, \;}
\label{eq:Sprod-avg}
\end{equation}
and that of \eqref{eq:eom5}-\eqref{eq:eom6}, $\langle \dot\zeta_x\rangle = \langle \dot\zeta_y\rangle = 0$, hence
\begin{equation}
\Big\langle \frac{p_i^2}{T_i} - 1 \Big\rangle = 0
\;\Rightarrow\; \boxed{\; \langle p_x^2 \rangle = T_x,\quad \langle p_y^2 \rangle = T_y. \;}
\label{eq:equipart}
\end{equation}
Long-time average of energy balance of Eq.~\eqref{eq:Edot-final}
\begin{equation}
\left\langle \dot E \right\rangle = - \left\langle \zeta_x p_x^2 + \zeta_y p_y^2 \right\rangle = 0.
 \label{eq:ss-energy}
\end{equation}
Using \eqref{eq:equipart} in \eqref{eq:ss-energy}:
\begin{equation}
\boxed{\; \langle T_x \rangle \langle \zeta_x\rangle + \langle T_x \rangle \langle \zeta_y\rangle = 0 \;}
\quad\Rightarrow\quad
\langle \zeta_x\rangle = -\frac{ \langle T_y \rangle}{ \langle T_x \rangle}\,\langle \zeta_y\rangle.
\label{eq:zx-from-zy}
\end{equation}
Using \eqref{eq:zx-from-zy} in \eqref{eq:Sprod-avg}
\begin{align}
\left\langle \dot S_{\mathrm{prod}} \right\rangle
&= \left\langle \zeta_x + \zeta_y \right\rangle
 = -\frac{ \langle T_y \rangle}{ \langle T_x \rangle}\,\langle \zeta_y\rangle + \langle \zeta_y\rangle \\
& = \langle \zeta_y\rangle \left(1 - \frac{ \langle T_y \rangle}{ \langle T_x \rangle}\right) \nonumber \\
&= \boxed{\; \langle \zeta_y\rangle \,\frac{ \langle T_x \rangle -  \langle T_y \rangle}{ \langle T_x \rangle}
= \langle \zeta_y\rangle\,\frac{\delta}{ \langle T_x \rangle}. \;}
\label{eq:final-entropy}
\end{align}

\paragraph{Phase-space divergence route\\\\}

Let $\Gamma = (x, y, p_x, p_y, \zeta_x, \zeta_y)$ and let $\dot{\Gamma}$ be the flow defined by Eqs.~\eqref{eq:eom1} - \eqref{eq:eom6}.  
The phase-space divergence (compressibility) is given by
\begin{equation}
\Lambda = \nabla_{\Gamma} \cdot \dot{\Gamma} 
= \sum_{a} \frac{\partial \dot{\Gamma}_a}{\partial \Gamma_a}.
\label{eq:compressibility}
\end{equation}
From Eqs.~\eqref{eq:eom1} - \eqref{eq:eom6} we have:
\[\frac{\partial \dot{x}}{\partial x} 
= \frac{\partial \dot{y}}{\partial y} 
= \frac{\partial \dot{\zeta}_x}{\partial \zeta_x} 
= \frac{\partial \dot{y}}{\partial \zeta_y} = 0,\]
\vskip -0.2cm
\[\frac{\partial \dot{p}_x}{\partial p_x} = -\zeta_x,
\quad\frac{\partial \dot{p}_y}{\partial p_y} = -\zeta_y.\]
Therefore:
\begin{equation}
\Lambda = -\zeta_x - \zeta_y.
\label{eq:lambda}
\end{equation}

Using \eqref{eq:Sprod-avg} we obtain the standard link between phase-space contraction and entropy production (with $k = 1$ units):
\vskip -0.4cm
\begin{equation}
\left\langle \dot{S}_{\mathrm{prod}} \right\rangle
= -\langle \Lambda \rangle.
\label{eq:entropy_link}
\end{equation}

Using the thermostat variables from our simulations, we tried calculating the $\Lambda$ and $\dot S_{prod}$ from formulations derived above. As showed in Table 2, $\Lambda$ values obtained from phase space contraction are in agreement with the sum of Lyapunov exponents (see Table 1). Consistently, they also agree with $\dot S_{prod}$.

\begin{table}[t]
\caption{Phase-space contraction rate using thermostat variables, $\zeta_x, \zeta_y $, for each thermostat pair in the anisotropic scan. Values are based on simulations with $10^{11}$ integration steps.}
\label{tab:Lambda-phase}
\centering
\begin{tabular}{cccccc}
\hline\hline
$\delta = 0.5 - T_y$ & $\langle \zeta_x \rangle$ & $\langle \zeta_y \rangle$ & $\Lambda = -\langle \zeta_x \rangle - \langle \zeta_y \rangle$ & $\dot S_{prod} = \frac{\langle \zeta_y \rangle \delta}{T_x}$\\
\hline
0.0   & 0.0001 & -0.0001 & 0.0        & 0.0    \\
0.05 & -0.0074 & 0.0083 & -0.0008 & 0.0008 \\
0.1   & -0.0146 & 0.0182 & -0.0036 & 0.0036 \\
0.15 & -0.0218 & 0.0311 & -0.0093 & 0.0093 \\
0.2   & -0.0291 & 0.0486 & -0.0194 & 0.0194 \\
0.25 & -0.0344 & 0.0687 & -0.0344 & 0.0344 \\
0.3   & -0.0345 & 0.0862 & -0.0517 & 0.0517 \\
0.35 & -0.0310 & 0.1034 & -0.0724 & 0.0724 \\
0.4   & -0.0241 & 0.1204 & -0.0963 & 0.0963 \\
0.45 & -0.0154 & 0.1541 & -0.1387 & 0.1387 \\
\hline\hline
\end{tabular}
\end{table}

\bibliographystyle{apsrev4-2}
\bibliography{references}

\begin{thebibliography}{14}%
\makeatletter
\providecommand \@ifxundefined [1]{%
 \@ifx{#1\undefined}
}%
\providecommand \@ifnum [1]{%
 \ifnum #1\expandafter \@firstoftwo
 \else \expandafter \@secondoftwo
 \fi
}%
\providecommand \@ifx [1]{%
 \ifx #1\expandafter \@firstoftwo
 \else \expandafter \@secondoftwo
 \fi
}%
\providecommand \natexlab [1]{#1}%
\providecommand \enquote  [1]{``#1''}%
\providecommand \bibnamefont  [1]{#1}%
\providecommand \bibfnamefont [1]{#1}%
\providecommand \citenamefont [1]{#1}%
\providecommand \href@noop [0]{\@secondoftwo}%
\providecommand \href [0]{\begingroup \@sanitize@url \@href}%
\providecommand \@href[1]{\@@startlink{#1}\@@href}%
\providecommand \@@href[1]{\endgroup#1\@@endlink}%
\providecommand \@sanitize@url [0]{\catcode `\\12\catcode `\$12\catcode
  `\&12\catcode `\#12\catcode `\^12\catcode `\_12\catcode `\%12\relax}%
\providecommand \@@startlink[1]{}%
\providecommand \@@endlink[0]{}%
\providecommand \url  [0]{\begingroup\@sanitize@url \@url }%
\providecommand \@url [1]{\endgroup\@href {#1}{\urlprefix }}%
\providecommand \urlprefix  [0]{URL }%
\providecommand \Eprint [0]{\href }%
\providecommand \doibase [0]{https://doi.org/}%
\providecommand \selectlanguage [0]{\@gobble}%
\providecommand \bibinfo  [0]{\@secondoftwo}%
\providecommand \bibfield  [0]{\@secondoftwo}%
\providecommand \translation [1]{[#1]}%
\providecommand \BibitemOpen [0]{}%
\providecommand \bibitemStop [0]{}%
\providecommand \bibitemNoStop [0]{.\EOS\space}%
\providecommand \EOS [0]{\spacefactor3000\relax}%
\providecommand \BibitemShut  [1]{\csname bibitem#1\endcsname}%
\let\auto@bib@innerbib\@empty
\bibitem [{\citenamefont {Hoover}(1985)}]{hoover1985canonical}%
  \BibitemOpen
  \bibfield  {author} {\bibinfo {author} {\bibfnamefont {W.~G.}\ \bibnamefont
  {Hoover}},\ }\href@noop {} {\bibfield  {journal} {\bibinfo  {journal}
  {Physical review A}\ }\textbf {\bibinfo {volume} {31}},\ \bibinfo {pages}
  {1695} (\bibinfo {year} {1985})}\BibitemShut {NoStop}%
\bibitem [{\citenamefont {Nos{\'e}}(1984)}]{nose1984molecular}%
  \BibitemOpen
  \bibfield  {author} {\bibinfo {author} {\bibfnamefont {S.}~\bibnamefont
  {Nos{\'e}}},\ }\href@noop {} {\bibfield  {journal} {\bibinfo  {journal}
  {Molecular Physics}\ }\textbf {\bibinfo {volume} {52}},\ \bibinfo {pages}
  {255} (\bibinfo {year} {1984})}\BibitemShut {NoStop}%
\bibitem [{\citenamefont {Hoover}\ and\ \citenamefont
  {Hoover}(2024)}]{hoover2024canonical}%
  \BibitemOpen
  \bibfield  {author} {\bibinfo {author} {\bibfnamefont {W.~G.}\ \bibnamefont
  {Hoover}}\ and\ \bibinfo {author} {\bibfnamefont {C.}~\bibnamefont
  {Hoover}},\ }\href@noop {} {\bibfield  {journal} {\bibinfo  {journal}
  {Computational Methods in Science \& Technology}\ }\textbf {\bibinfo {volume}
  {30}} (\bibinfo {year} {2024})}\BibitemShut {NoStop}%
\bibitem [{\citenamefont {Evans}\ and\ \citenamefont
  {Morriss}(2008)}]{evans2008statistical}%
  \BibitemOpen
  \bibfield  {author} {\bibinfo {author} {\bibfnamefont {D.~J.}\ \bibnamefont
  {Evans}}\ and\ \bibinfo {author} {\bibfnamefont {G.}~\bibnamefont
  {Morriss}},\ }\href@noop {} {\emph {\bibinfo {title} {Statistical Mechanics
  of Nonequilibrium Liquids}}}\ (\bibinfo  {publisher} {Cambridge University
  Press},\ \bibinfo {year} {2008})\BibitemShut {NoStop}%
\bibitem [{\citenamefont {Martyna}\ \emph {et~al.}(1992)\citenamefont
  {Martyna}, \citenamefont {Klein},\ and\ \citenamefont
  {Tuckerman}}]{martyna1992nose}%
  \BibitemOpen
  \bibfield  {author} {\bibinfo {author} {\bibfnamefont {G.~J.}\ \bibnamefont
  {Martyna}}, \bibinfo {author} {\bibfnamefont {M.~L.}\ \bibnamefont {Klein}},\
  and\ \bibinfo {author} {\bibfnamefont {M.}~\bibnamefont {Tuckerman}},\
  }\href@noop {} {\bibfield  {journal} {\bibinfo  {journal} {The Journal of
  chemical physics}\ }\textbf {\bibinfo {volume} {97}},\ \bibinfo {pages}
  {2635} (\bibinfo {year} {1992})}\BibitemShut {NoStop}%
\bibitem [{\citenamefont {Hoover}\ \emph {et~al.}(1991)\citenamefont {Hoover},
  \citenamefont {Craig}, \citenamefont {Posch}, \citenamefont {Holian},\ and\
  \citenamefont {Hoover}}]{hoover1991heat}%
  \BibitemOpen
  \bibfield  {author} {\bibinfo {author} {\bibfnamefont {W.~G.}\ \bibnamefont
  {Hoover}}, \bibinfo {author} {\bibfnamefont {E.}~\bibnamefont {Craig}},
  \bibinfo {author} {\bibfnamefont {H.~A.}\ \bibnamefont {Posch}}, \bibinfo
  {author} {\bibfnamefont {B.~L.}\ \bibnamefont {Holian}},\ and\ \bibinfo
  {author} {\bibfnamefont {C.~G.}\ \bibnamefont {Hoover}},\ }\href@noop {}
  {\bibfield  {journal} {\bibinfo  {journal} {Chaos An Interdisciplinary
  Journal of Nonlinear Science}\ }\textbf {\bibinfo {volume} {1}},\ \bibinfo
  {pages} {343} (\bibinfo {year} {1991})}\BibitemShut {NoStop}%
\bibitem [{\citenamefont {Squires}\ and\ \citenamefont
  {Quake}(2005)}]{squires2005microfluidics}%
  \BibitemOpen
  \bibfield  {author} {\bibinfo {author} {\bibfnamefont {T.~M.}\ \bibnamefont
  {Squires}}\ and\ \bibinfo {author} {\bibfnamefont {S.~R.}\ \bibnamefont
  {Quake}},\ }\href@noop {} {\bibfield  {journal} {\bibinfo  {journal} {Reviews
  of modern physics}\ }\textbf {\bibinfo {volume} {77}},\ \bibinfo {pages}
  {977} (\bibinfo {year} {2005})}\BibitemShut {NoStop}%
\bibitem [{\citenamefont {Cahill}\ \emph {et~al.}(2014)\citenamefont {Cahill},
  \citenamefont {Braun}, \citenamefont {Chen}, \citenamefont {Clarke},
  \citenamefont {Fan}, \citenamefont {Goodson}, \citenamefont {Keblinski},
  \citenamefont {King}, \citenamefont {Mahan}, \citenamefont {Majumdar} \emph
  {et~al.}}]{cahill2014nanoscale}%
  \BibitemOpen
  \bibfield  {author} {\bibinfo {author} {\bibfnamefont {D.~G.}\ \bibnamefont
  {Cahill}}, \bibinfo {author} {\bibfnamefont {P.~V.}\ \bibnamefont {Braun}},
  \bibinfo {author} {\bibfnamefont {G.}~\bibnamefont {Chen}}, \bibinfo {author}
  {\bibfnamefont {D.~R.}\ \bibnamefont {Clarke}}, \bibinfo {author}
  {\bibfnamefont {S.}~\bibnamefont {Fan}}, \bibinfo {author} {\bibfnamefont
  {K.~E.}\ \bibnamefont {Goodson}}, \bibinfo {author} {\bibfnamefont
  {P.}~\bibnamefont {Keblinski}}, \bibinfo {author} {\bibfnamefont {W.~P.}\
  \bibnamefont {King}}, \bibinfo {author} {\bibfnamefont {G.~D.}\ \bibnamefont
  {Mahan}}, \bibinfo {author} {\bibfnamefont {A.}~\bibnamefont {Majumdar}},
  \emph {et~al.},\ }\href@noop {} {\bibfield  {journal} {\bibinfo  {journal}
  {Applied physics reviews}\ }\textbf {\bibinfo {volume} {1}} (\bibinfo {year}
  {2014})}\BibitemShut {NoStop}%
\bibitem [{\citenamefont {Hoover}\ and\ \citenamefont
  {Hoover}(2015)}]{hoover2015comparison}%
  \BibitemOpen
  \bibfield  {author} {\bibinfo {author} {\bibfnamefont {W.}~\bibnamefont
  {Hoover}}\ and\ \bibinfo {author} {\bibfnamefont {C.}~\bibnamefont
  {Hoover}},\ }\href@noop {} {\bibfield  {journal} {\bibinfo  {journal}
  {Computational Methods in Science and Technology}\ }\textbf {\bibinfo
  {volume} {21}} (\bibinfo {year} {2015})}\BibitemShut {NoStop}%
\bibitem [{\citenamefont {Evans}\ \emph {et~al.}(1993)\citenamefont {Evans},
  \citenamefont {Cohen},\ and\ \citenamefont {Morriss}}]{evans1993probability}%
  \BibitemOpen
  \bibfield  {author} {\bibinfo {author} {\bibfnamefont {D.~J.}\ \bibnamefont
  {Evans}}, \bibinfo {author} {\bibfnamefont {E.~G.~D.}\ \bibnamefont
  {Cohen}},\ and\ \bibinfo {author} {\bibfnamefont {G.~P.}\ \bibnamefont
  {Morriss}},\ }\href@noop {} {\bibfield  {journal} {\bibinfo  {journal}
  {Physical review letters}\ }\textbf {\bibinfo {volume} {71}},\ \bibinfo
  {pages} {2401} (\bibinfo {year} {1993})}\BibitemShut {NoStop}%
\bibitem [{\citenamefont {Benettin}\ \emph {et~al.}(1976)\citenamefont
  {Benettin}, \citenamefont {Galgani},\ and\ \citenamefont
  {Strelcyn}}]{benettin1976kolmogorov}%
  \BibitemOpen
  \bibfield  {author} {\bibinfo {author} {\bibfnamefont {G.}~\bibnamefont
  {Benettin}}, \bibinfo {author} {\bibfnamefont {L.}~\bibnamefont {Galgani}},\
  and\ \bibinfo {author} {\bibfnamefont {J.-M.}\ \bibnamefont {Strelcyn}},\
  }\href@noop {} {\bibfield  {journal} {\bibinfo  {journal} {Physical Review
  A}\ }\textbf {\bibinfo {volume} {14}},\ \bibinfo {pages} {2338} (\bibinfo
  {year} {1976})}\BibitemShut {NoStop}%
\bibitem [{\citenamefont {Posch}\ and\ \citenamefont
  {Hoover}(1989)}]{posch1989equilibrium}%
  \BibitemOpen
  \bibfield  {author} {\bibinfo {author} {\bibfnamefont {H.~A.}\ \bibnamefont
  {Posch}}\ and\ \bibinfo {author} {\bibfnamefont {W.~G.}\ \bibnamefont
  {Hoover}},\ }\href@noop {} {\bibfield  {journal} {\bibinfo  {journal}
  {Physical Review A}\ }\textbf {\bibinfo {volume} {39}},\ \bibinfo {pages}
  {2175} (\bibinfo {year} {1989})}\BibitemShut {NoStop}%
\bibitem [{\citenamefont {B{\'a}lint}\ \emph {et~al.}(2023)\citenamefont
  {B{\'a}lint}, \citenamefont {Bruin},\ and\ \citenamefont
  {Terhesiu}}]{balint2023periodic}%
  \BibitemOpen
  \bibfield  {author} {\bibinfo {author} {\bibfnamefont {P.}~\bibnamefont
  {B{\'a}lint}}, \bibinfo {author} {\bibfnamefont {H.}~\bibnamefont {Bruin}},\
  and\ \bibinfo {author} {\bibfnamefont {D.}~\bibnamefont {Terhesiu}},\
  }\href@noop {} {\bibfield  {journal} {\bibinfo  {journal} {Probability Theory
  and Related Fields}\ }\textbf {\bibinfo {volume} {186}},\ \bibinfo {pages}
  {159} (\bibinfo {year} {2023})}\BibitemShut {NoStop}%
\bibitem [{\citenamefont {Holian}\ \emph {et~al.}(1995)\citenamefont {Holian},
  \citenamefont {Voter},\ and\ \citenamefont
  {Ravelo}}]{holian1995thermostatted}%
  \BibitemOpen
  \bibfield  {author} {\bibinfo {author} {\bibfnamefont {B.~L.}\ \bibnamefont
  {Holian}}, \bibinfo {author} {\bibfnamefont {A.~F.}\ \bibnamefont {Voter}},\
  and\ \bibinfo {author} {\bibfnamefont {R.}~\bibnamefont {Ravelo}},\
  }\href@noop {} {\bibfield  {journal} {\bibinfo  {journal} {Physical Review
  E}\ }\textbf {\bibinfo {volume} {52}},\ \bibinfo {pages} {2338} (\bibinfo
  {year} {1995})}\BibitemShut {NoStop}%
\end{thebibliography}%

\end{document}